\documentclass[aps,twocolumn]{revtex4}
\usepackage [dvips]{graphicx}
\def\beq{\begin{equation}}
\def\eeq{\end{equation}}
\begin{document}
\title{\bf
 Nuclear Superfluidity and Specific Heat in the Inner Crust
 of Neutron Stars}

\author{\rm
Nicolae Sandulescu}

\bigskip

\address {\rm
Institute of Physics and Nuclear Engineering, 76900 Bucharest, 
  Romania\\
  Institut de Physique Nucl\'eaire, Universit\'e Paris-Sud,
           F-91406 Orsay Cedex, France \\
  Royal Institute of Technology, Alba Nova,
            SE-10691, Stockholm, Sweden }

\begin{abstract}
 We analyse the temperature dependence of pairing correlations
 in the inner crust matter of neutron stars. The study is done 
 in a finite-temperature HFB approach and by using a zero range
 pairing force adjusted to the pairing  properties of infinite
 neutron matter. Within the same approach we investigate how the
 specific heat of the inner crust  
 depends on temperature,  matter inhomogeneity, and the assumption
 used for the pairing force.
 It is shown that in a physical relevant range of densities the pairing
 properties of inner crust matter depend significantly on temperature.
 The finite-temperature HFB calculations  show also that the specific 
 heat is rather sensitive to the presence of nuclear clusters
 inside the inner crust. However, the most dramatic change of
 the specific  heat is determined  by the scenario used for 
 the neutron matter  superfluidity.

\end{abstract}

\pacs{PACS number(s): 25.70.Ef,23.50.+z,25.60+v,21.60.Cs}

\maketitle

\newpage

 The inner crust of neutron stars consists of a lattice of
 neutron-rich nuclei immersed in a sea of unbound neutrons
 and relativistic electrons. Down to the inner edge of the
 crust, the crystal lattice is most probably formed by 
 spherical nuclei. More inside the star, before the 
 nuclei dissolve completely into the liquid of the core, 
 the nuclear matter can develop other exotic configurations
 as well, i.e, rods,  plates, tubes, and bubbles (for a discussion
 of inner crust matter see \cite{pethick} and  references therein).
 The thickness of the inner crust is rather small, of the order
 of one kilometer, and its mass is only about  1$\%$ of the 
 neutron star mass. However, in spite of its small size, the properties
 of inner crust matter, especially its superfluid  properties, 
 have important consequences for the dynamics  and the thermodynamics
 of neutron stars. 

 The superfluid properties of the inner crust have been 
 considered long ago in connection with post-glitch timing
 observations and colling processes \cite{pines,saul,cooling}. However, 
 although the  neutron star matter  superfluidity has been intensively 
 studied in the last decades \cite{lombardo}, so far only a  few 
 microscopic calculations have been done for the  superfluidity of inner 
 crust matter.
  The existing calculations are  done in the framework of the
 Hartree-Fock-Bogoliubov (HFB) approach and at zero temperature 
 \cite{broglia,barranco,pizzochero, sandulescu}. As it is well-known,
 the limitation to zero temperature is justified only if the pairing
 gap of the superfluid is much larger than its temperature.
 This is not generally the case for the inner crust matter, 
 where this condition is fulfiled only for some limited
 density regions.

 The purpose of this paper is to study how the pairing 
 properties of inner crust matter are changing with the
 temperature and what are the  consequences of these changes
 on the specific heat. The present investigation is done in a 
 finite-temperature HFB approach. In the calculations
 both the mean field and the pairing field are determined 
 self-consistently by using a Skyrme-type interaction and an
 effective pairing force adjusted to the superfluid properties
 of infinite neutron matter. The setting of the calculations
 follows Ref.\cite{sandulescu}, where the same two-body forces
 have been used for describing the pairing properties of inner
 crust matter at zero temperature.
 
 The finite-temperature HFB (FT-HFB) equations have a structure
 similar to the HFB equations at zero temperature \cite{goodman}.
 For zero range forces and spherical symmetry the radial FT-HFB
 equations can be written as:
\beq
\begin{array}{c}
\left( \begin{array}{cc}
h_T(r) - \lambda & \Delta_T (r) \\
\Delta_T (r) & -h_T(r) + \lambda 
\end{array} \right)
\left( \begin{array}{c} \mathfrak{U}_i (r) \\
 \mathfrak{V}_i (r) \end{array} \right) = E_i
\left( \begin{array}{c} \mathfrak{U}_i (r) \\
 \mathfrak{V}_i (r) \end{array} \right) ~,
\end{array}
\label{1}
\eeq
where $E_i$ is the quasiparticle energy, $U_i$, $V_i$ are 
the components of the radial FT-HFB wave function
and $\lambda$ is the chemical potential.
The quantity $h_T(r)$ is the thermal averaged mean field 
hamiltonian and $\Delta_T (r)$ is the thermal averaged
pairing field. 
In a self-consistent calculation based on a 
Skyrme-type force 
as used in this study, $h_T(r)$ is expressed in terms of thermal
averaged densities, i.e., kinetic energy density $\tau_T(r)$, particle 
density $\rho_T(r)$ and spin density $J_T(r)$, in the same way as in 
the Skyrme-HF approach \cite{vautherin}. The thermal averaged densities
are given by:
\beq
\rho_T(r) =\frac{1}{4\pi} \sum_{i} (2j_i+1) [ \mathfrak{V}_i^* (r) 
\mathfrak{V}_i (r) (1 - f_i ) \\
+ \mathfrak{U}_i^* (r) \mathfrak{U}_i (r) f_i ] 
\label{2}
\eeq
\begin{eqnarray}
J_T(r) & = & \frac{1}{4\pi} \sum_i (2j_i+1) 
[j_i(j_i+1)-l_i(l_i+1)-\frac{3}{4}] \nonumber  \\
& & \times \{ V_i^2 (1-f_i) + U_i^2 f_i \}
\label{4}
\end{eqnarray}
\begin{eqnarray}
\tau_T(r) & = & \frac{1}{4\pi} \sum_{i} (2j_i+1)
\{ [(\frac{dV_i}{dr}-\frac{V_i}{r})^2 +\frac{l_i(l_i+1)}{r^2} V_i^2 ]
\nonumber \\
& & \times (1 - f_i)
+ [(\frac{dU_i}{dr}-\frac{U_i}{r})^2 +\frac{l_i(l_i+1)}{r^2}
U_i^2]f_i \} , \nonumber \\
\end{eqnarray}
where $f_i = [1 + exp ( E_i/k_B T)]^{-1}$ is the 
Fermi distribution, $k_B$ is the Boltzmann constant and T is the 
temperature. The summations in the equations above are over the
whole positive energy quasiparticle spectrum, including the
continuum states. In the latter case the summations should be
 replaced by an integral over the energy \cite{grasso}.

The thermal averaged pairing field is calculated with a  density 
dependent contact force of the following form \cite{bertsch}:
\beq
V (\mathbf{r}-\mathbf{r^\prime}) = V_0 [1 -\eta 
(\frac{\rho(r)}{\rho_0})^{\alpha}] 
\delta(\mathbf{r}-\mathbf{r^\prime}) 
\equiv V_{eff}(\rho(r)) \delta(\mathbf{r}-\mathbf{r^\prime}) ,
\eeq  
where $\rho(r)$ is the baryonic density.
 With this force the thermal averaged pairing field is local
and is given by:

\begin{eqnarray}
\Delta_T(r) & = & V_{eff}(\rho(r)){
\frac{1}{4\pi} \sum_{i} (2j_i+1) \mathfrak{U}_i^* (r) 
\mathfrak{V}_i (r)
(1 - 2f_i )} \nonumber \\
& \equiv & V_{eff}(\rho(r)) \kappa_T (r) ,
\end{eqnarray}
where $\kappa_T(r)$ is the thermal averaged pairing tensor.
Due to the density dependence of the pairing force, the
thermal averaged mean field hamiltonian $h_T(r)$ depends 
also on $\kappa_T$.

The inner crust matter is usually divided into independent
Wigner-Seitz cells, each cell containing in its center
a neutron-rich nucleus surrounded by unbound neutrons and 
permeated by a relativistic electron gas uniformly distributed
inside the cell. In this study the FT-HFB equations are solved
for some representative spherical cells 
determined in Ref.\cite{negele}.
To generate far from the nucleus a constant density
corresponding to the neutron gas, the FT-HFB equations are 
solved by imposing Dirichlet-Neumann boundary conditions
at the edge of the cell \cite{negele}, i.e., all wave
functions of even parity vanish and the derivatives of 
odd-parity wave functions vanish.

 In the FT-HFB calculations we use for the particle-hole channel
 the Skyrme effective interaction SLy4 \cite{sly4}, which 
 has been adjusted to describe properly the mean field properties
 of neutron-rich nuclei and infinite neutron matter. For the
 effective pairing force the choice is  more  problematic. 
 A realistic FT-HFB calculation of crust superfluidity should be
 based on a pairing force able to describe reasonably well at least
 the pairing properties of neutron matter, which forms the main part
 of the inner crust. However, the magnitude of pairing correlations 
 in neutron matter is still a subject of debate. On one hand, 
 BCS calculations based on a Gogny force commonly used 
 in finite nuclei \cite{gogny}, give for the pairing 
 gap of infinite neutron matter a maximum value of  about 
 3.2 MeV at a Fermi momentum $k_F \approx 0.9$ fm$^{-1}$
 \cite{shen}. 
 On the other hand, the microscopic calculations based on 
 induced  interactions predict for the maximum value of the
 gap much smaller values, of around 1 MeV \cite{shen,wambach,brown}.
 These two different scenarios for the neutron  matter superfluidity
 are used here to determine the parameters of the pairing
 force (5) employed in the FT-HFB  calculations.
 Thus, for each Wigner-Seitz cell we perform two FT-HFB calculations,
 with two sets of parameters for the pairing force. For the first
 calculation we use the parameters: $V_{0}$=-430 MeV  fm$^3$, 
 $\eta$=0.7, and $\alpha$=0.45. With these parameters and
 with a  cut-off energy for the quasiparticle spectrum equal
 to 60 MeV one obtains approximately the pairing gap
 given by the Gogny force in nuclear matter \cite{bertsch,garrido}.
 In the second calculation we reduce the strength of the force
 to the value $V_0$=-330 MeV fm$^3$. With this value of the 
 strength we simulate the second scenario for the neutron
 matter superfluidity, in which the maximum
 gap in neutron matter is around 1 MeV. In principle, the other 
 two parameters of the pairing force should be adjusted to the 
 density dependence of the gap in neutron matter corresponding 
 to the calculations based on induced interactions. 
 However, since at present this dependence is not 
 well-established, we keep for the parameters $\eta$ and $\alpha$
 the same values as in the first calculation.

 The FT-HFB results are analysed here for two representative 
 Wigner-Seitz cells chosen from Ref.\cite{negele}.
 These cells contain Z=50 protons and have rather different
 baryonic densities, i.e., 0.0204 fm$^{-3}$ and  0.00373 fm$^{-3}$. 
 The cells, which contain N=1750 and N=900 neutrons, respectively,
 are denoted below as a nucleus with Z protons and N neutrons,
 i.e., $^{1800}$Sn and $^{950}$Sn. For the baryonic densities
 and the nucleonic numbers listed above, the radii of the cells
 are equal to 27.6 fm and 39.3 fm, respectively. In the calculations
 we integrate the FT-HFB  equations up to 27.6 fm  for both cells.
 Up to this distance in the second cell there are N=279 neutrons.
 Since at 27.6 fm we are already far out in the region of 
 the neutron gas, the pairing field of the second cell
 is not influenced by this choice of the integration radius.

 The FT-HFB calculations are done up to a maximum temperature of
 T=0.5 MeV, which is covering the temperature range of physical
 interest. According to numerical simulations \cite{riper}, 
 the temperature in the inner crust is increasing from  about 
 T=0.1 MeV in the low-density region to about T=0.3 MeV at the
 inner edge of the crust. This temperature profile corresponds to
 the state of matter before the cooling wave is passing through the 
 inner crust from the core region to the star surface. 

\begin{figure}[h]
\begin{center}
\includegraphics*[scale=0.35,angle=-90.]{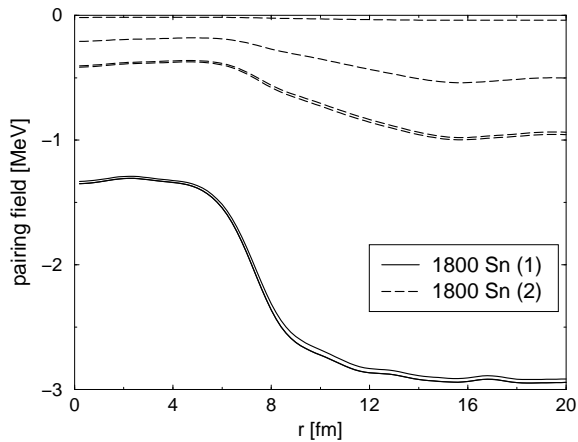}
\caption{
Neutron pairing fields for the cell $^{1800}$Sn
calculated at various temperatures. The numbers 1 and 2 
which follow the cell symbol (see the inset) indicate 
the variant of the pairing force used in the calculations. 
The full and the dashed lines corresponds (from bottom upwards)
to the set of temperatures T=$\{0.0, 0.5\}$
MeV and T=$\{0.0, 0.1, 0.3, 0.5\}$MeV, respectively.}
\end{center}
\end{figure}

 The temperature dependence of the pairing fields in the two cells
 presented above is shown in Figs.1-2. First, we notice that for
 all temperatures the nuclear clusters modify significantly the
 profile of the pairing field.  Thus, for the high density cell
 $^{1800}$Sn we can see that the pairing field is  strongly
 decreasing  (in absolute value) in the cluster region.
 However, as already noticed in zero- temperature HFB calculations 
 \cite{sandulescu}, the situation is completely different for the 
 low-density cell $^{950}$Sn, where the pairing field is strongly
 increasing in the  transition region from the neutron gas to the
 nuclear cluster. These different behaviours of the pairing fields
 are connected to the density dependence of the pairing gap in
 neutron matter \cite{shen,wambach,brown}.

\begin{figure}[h]
\begin{center}
\includegraphics*[scale=0.35,angle=-90.]{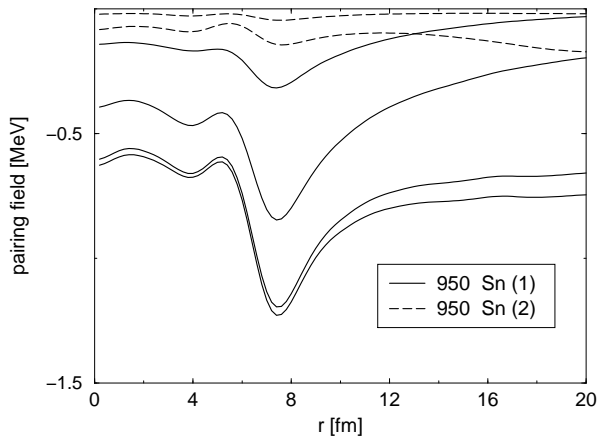}
\caption{The same as in Fig.1, but for the cell $^{950}$Sn.
The full and the dashed lines corresponds (from bottom 
upwards) to the set of temperatures T=$\{0.0, 0.1, 0.3, 0.5\}$
and T=$\{0.0, 0.1\}$MeV, respectively}
\end{center}
\end{figure}

 As expected, the dependence on temperature of the pairing field 
 is weak when the gap values are much larger than the  temperature.
 This is the case for the high-density cell $^{1800}$Sn  and the
 first pairing force. For all other cases the temperature 
 dependence of the pairing field is significant. This is
 clearly seen in the low-density cell $^{950}$Sn. From Fig.1
 we can see that for the second pairing force even in the
 high-density cell $^{1800}$Sn the pairing field is changing
 significantly  with the temperature.

 To analyse how the pairing correlations are globally changing
 in the cell as a function  of temperature, in Fig.3 we plot
 the  pairing energy per neutron for the cases in which its
 temperature dependence
 is significant. In Fig.3 are shown also the results obtained  when
 in the same cell are distributed uniformly only the neutrons.
 One can notice that the pairing energy of the non-uniform system,
 i.e., cluster plus the neutron gas, has a rather different  temperature
 dependence  compared to the uniform neutron matter. From Fig.3 we can
 eventually extract a critical temperature for the superfluid-normal
 phase 
 transition in the non-uniform  nuclear matter of the cell. However, as
 seen for the cell $^{950}$Sn, we should be aware of the fact that close
 to this averaged critical temperature the non-uniform system may still
 have regions in which the pairing field has non-negligible values.

\begin{figure}[h]
\begin{center}
\includegraphics*[scale=0.35,angle=-90.]{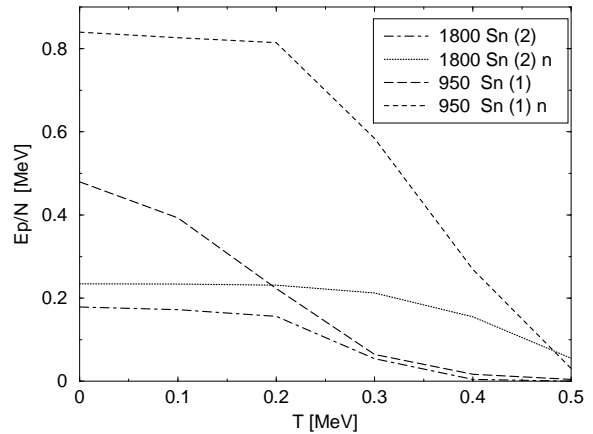}
\caption{
Pairing energy per neutron as a function of temperature.
The calculated values, done in step of $\Delta$T=0.1 MeV,
are joined by straight lines. The symbol "n" following the
cell symbol (see the inset) refers to the results obtained 
when in the 
corresponding cell are distributed uniformly only the neutrons. 
The rest of the notations are the same as in Fig.1.}
\end{center}
\end{figure}

 We turn now to the specific heat provided by the FT-HFB calculations.
 The specific heat of a given cell of volume V is defined by:

\beq
C_V = \frac{1}{V}\frac{\partial \mathcal{E}(T)}{\partial T} ,
\eeq
where $\mathcal{E}(T)$ is the total energy 
of the baryonic matter inside the cell, i.e.,

\beq
\mathcal{E}(T) = \sum_i f_i E_i .
\eeq
 Due to the energy gap in the excitation spectrum, the specific
 heat of a superfluid system is dramatically  reduced  compared
 to its value in the normal phase. Since the specific heat depends
 exponentially on the energy gap, its value for a Wigner-Seitz cell
 is very sensitive to the local variations of the pairing field
 induced by the nuclear clusters. This can be clearly
 seen in Fig.4, where the specific heat is plotted for the cell 
 $^{1800}$Sn and for the neutrons uniformly distributed
 in the same cell. One can notice that at T=0.1 MeV and for
 the first pairing force the presence of the cluster increases
 the specific heat by about 6 times  compared to the value for
 the uniform neutron gas. A comparable effect of the nuclear
 cluster on the specific heat of the cell $^{1800}$Sn was
 found earlier by using in Eq.(8) the quasiparticle spectrum
 provided by a zero-temperature HFB calculation
 \cite{barranco,pizzochero}.
 This approximation is justified only when the temperature
 variation of the pairing field in the Wigner-Seitz cells 
 is small, as in the case of high density cells and a Gogny
 type force. 

\begin{figure}[h]
\begin{center}
\includegraphics*[scale=0.35,angle=-90.]{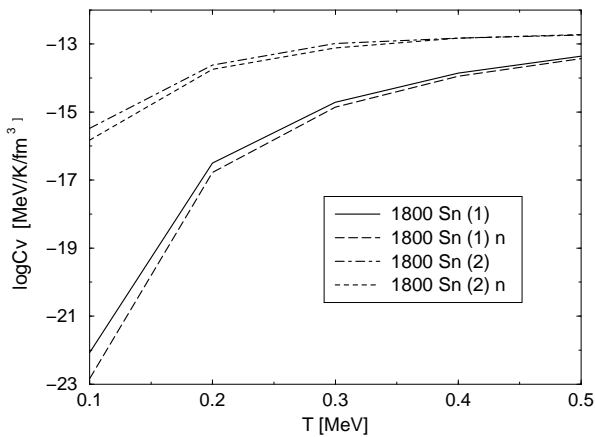}
\caption{
Specific heat for the cell $^{1800}$Sn as a function of
temperature. The notations used in the inset and the
representation of the calculated values are the same
as in Figs.1-3.}
\end{center}
\end{figure}

\begin{figure}[h]
\begin{center}
\includegraphics*[scale=0.35,angle=-90.]{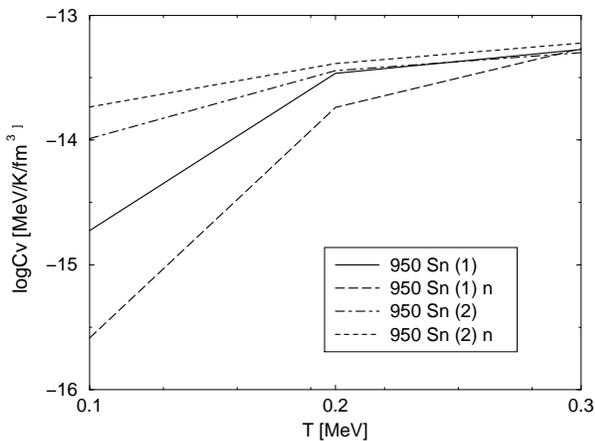}
\caption{ The same as in Fig.4, but for the cell $^{950}$Sn.}
\end{center}
\end{figure}

 The most striking fact seen in Fig.4 is the huge difference
 between the predictions of the two pairing forces. Thus,
 for T=0.1 MeV this difference amounts to about 7 orders
 of magnitude. This discrepancy between the predicted values
 of specific heat, generated by the changes in the pairing
 gaps, is decreasing rapidly with the temperature, but it
 remains significant up to  T=0.5 MeV. From Fig.4 we can
 see also that for the cell $^{1800}$Sn the effect of the
 cluster on specific heat is similar for both pairing forces.  

 The behaviour of the specific heat for the low- density cell
 $^{950}$Sn is shown in Fig.5. For the first pairing force
 we can also see that at T=0.1 MeV the cluster increases
 the specific heat by about the same factor as in the
 cell $^{1800}$Sn. However, for the second pairing force
 the situation is opposite: the presence of the nucleus 
 decreases the specific heat instead of increasing it. 
 From Fig.5 we can also see that
 the specific heat predicted by the two pairing forces are
 not very much different (notice the expanded scale) for
 this low-density cell. Due to the fast decreasing of the
 pairing correlations with the temperature, at T=0.3 MeV
 all calculations provide almost the same specific heat.

 We have also  performed FT-HFB calculations for two other cells 
 with 40 protons and having the densities equal to 
 0.0475 fm$^{-3}$ and 0.00159 fm$^{-3}$. We found that
 the temperature dependence  of the pairing field and specific
 heat in the first and in the second cell with Z=40 
 is very similar to the one in the cells $^{1800}$Sn and 
 $^{950}$Sn, respectively.

In conclusion, we have shown that the superfluid 
properties and the specific heat of inner crust matter are
affected rather strongly and in a non-trivial way by the
temperature of the crust. Therefore, the temperature dependence
should be included in the microscopic calculations aiming at 
describing accurately the superfluid properties of inner crust matter.
However, the most serious problem in microscopic calculations
remains the pairing interaction. As seen above, shifting to 
the second pairing force adjusted to the nuclear matter properties 
calculated with induced interactions, the specific heat of high 
density cells increases by several orders of magnitude. 
Consequently, for the second pairing force the neutron and
the electron contributions to the total specific heat of
high-density cells are becoming competitive, at variance with 
the results based on a Gogny force scenario \cite{pizzochero}. 
To analyse the implications of these two different scenarios for the
pairing force on physical properties as the cooling time one needs
to calculate within the FT-HFB approach the heat diffusion along the
whole inner crust. The succesive step would be to include also into the
microscopic calculations the ground state correlations of the inner
crust matter, which can modify significantly the neutron single-particle 
properties related to the heat diffusion (e.g., effective mass 
and the mean free path) \cite{wei}. It is to be expected that such
microscopic calculations could be eventually used to determine
accurately the consequences of inner crust superfluidity on the 
cooling time of neutron stars.

\end{document}